\documentclass[conference]{IEEEtran}
\IEEEoverridecommandlockouts

\usepackage{cite}
\usepackage{amsmath,amssymb,amsfonts}
\usepackage{graphicx}
\usepackage{textcomp}
\usepackage{xcolor}
\usepackage{amsmath}
\usepackage{enumitem,kantlipsum}

\usepackage{dirtytalk}

\usepackage{amsmath}
\usepackage{algorithm}
\usepackage{algpseudocode}
\usepackage{amsfonts}
\usepackage{color}
\usepackage{tcolorbox}
\usepackage{amsthm}

\newtheorem{theorem}{Theorem}

\newtheorem{lemma}{Lemma}

\newtheorem{prop}{Proposition}

\theoremstyle{definition}

\newtheorem{definition}{Definition}

\usepackage[table]{xcolor}
\newcommand{\off}[1]{}
\def\IsDraft{} 

\usepackage{authblk}

\ifdefined\IsDraft
    \newcommand{\authnote}[2]{{\bf [{\color{red} #1's Note:} {\color{blue} #2}]}}
\else
    \newcommand{\authnote}[2]{}
\fi

\def\BibTeX{{\rm B\kern-.05em{\sc i\kern-.025em b}\kern-.08em
    T\kern-.1667em\lower.7ex\hbox{E}\kern-.125emX}}

\title{Scalable Multiterminal Key Agreement via Error-Correcting Codes}

\author[1]{
Benjamin D. Kim
}

\author[2]{
Daniel Alabi
}

\author[3]{
Lav R. Varshney
}

\affil[1,2,3] {
University of Illinois Urbana-Champaign
}

\affil[3] {
Stony Brook University
}

\begin{document}

\maketitle

\begin{abstract}
     We explore connections between secret sharing and secret key agreement, which yield a simple and scalable multiterminal key agreement protocol. In our construction, we use error-correcting codes, specifically Reed-Solomon codes with threshold reconstruction, to ensure no information is leaked to an eavesdropper. We then derive novel bounds for both full-rank maximum distance separable codes and our scheme's secret key capacity, using key capacity's duality with multivariate mutual information.
\end{abstract}

\section{Introduction}
The problem where multiple parties wish to securely establish a cryptographic key has long been studied in cryptology. Advancements in classical cryptography enable secure key distribution via public-key cryptography (PKC), where one party encrypts a secret with a public key and only a user with the secret key is able to decrypt \cite{10.1145/359340.359342}. Classical cryptography assumes adversaries have limited computational power, rendering techniques such as the prime factorization problem \cite{riesel1994prime} crucial for designing cryptographic protocols. More recently, in preparation for the quantum computing era, protocols are using lattice-based primitives. The resulting schemes resist quantum attacks for which no sub-exponential complexity quantum attacks are known \cite{nejatollahi2019post, bos2018crystals, shor1999polynomial}. 

\begin{figure}[!t]
    \centering
    \includegraphics[width=0.55\columnwidth]{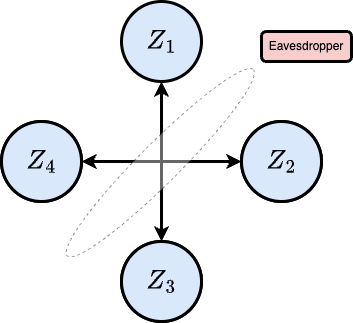}
    \caption{Multiple terminals $\{Z_1, Z_2, Z_3, Z_4 \}$ communicating with one another in public discussion. An eavesdropper listens in to their communication. In this work, their goal is to establish the same secret key without giving any information about the key to the eavesdropper.}
    \label{fig:overview}
\end{figure}

In contrast,
information-theoretic security considers an adversary with unlimited computational power. A similar key distribution problem in information-theoretic security is \textbf{secret key agreement} (SKA), where parties begin with correlated private sources and use those sources along with public discussion (see Fig.~\ref{fig:overview}) to establish a secret key. In this setting, the fundamental limit is measured by secret key capacity, the maximum achievable rate under discussion with an eavesdropper listening. 

Another commonly studied problem in information-theoretic security is secret sharing, first proposed by Shamir and Blakely \cite{shamir1979share}. In this problem, a dealer encodes a secret into $n$ shares such that $ k \le n$ shares suffice to reconstruct the secret, while every set of shares fewer than $k$ leaks no information about the secret, in that there is no mutual information between the two. Error-correcting codes also have a similar property: a message is encoded into a block of length $n$, and at least $k$ symbols from the block are typically required to fully decode back to the original message \cite{peterson1972error, huffman2010fundamentals}. Maximum distance separable (MDS) codes, and more specifically, Reed-Solomon codes are commonly used as error-correcting codes and can also serve as the basis for secret sharing schemes \cite{mceliece1981sharing}. 

In this work, we use linear error-correcting codes to explore the duality between secret sharing and secret key agreement, using MDS codes to develop a novel secret key agreement protocol. Through our scheme, we present an interesting characterization of MDS codes for SKA. Our key contributions are as follows.

\begin{itemize}
    \item To the best of our knowledge, we are the first to explore the duality between secret sharing and secret key agreement, utilizing their fundamental properties to present a simple and scalable MDS code-based SKA scheme. 
    \item We prove security of the scheme, derive new capacity bounds, and provide a multivariate mutual information (MMI) based analysis tailored to this construction. Compared to Reing et al.\ \cite{reing2019maximizing}, who analyze error-correcting codes that generally maximize MMI, our analysis focuses on calculating existing MMI measures specifically for our scheme.
\end{itemize}
The rest of the paper is organized as follows. Sec.~2 summarizes related work. Sec.~3 presents our coding-based SKA scheme. Sec.~4 derives key capacity and MMI results. Lastly, Secs.~5 and 6 discuss our results and conclude.

\section{Related Work}
SKA protocols were originally introduced by Maurer and Wolf \cite{maurer1996towards} and by Ahlswede and Csisz\'ar  \cite{csiszar2004secrecy}. In these protocols, multiple communication terminals are given correlated private sources and publicly communicate to establish a private key. Compared to traditional key exchange protocols that assume a computationally-bounded adversary, SKA protocols achieve an information-theoretic security guarantee, secure against adversaries with unbounded computational power. SKA has several similar properties relating to the problem of distributed source coding (DSC) \cite{wyner2003recent, pradhan2000distributed}, specifically Slepian-Wolf coding \cite{slepian2003noiseless}, where users start with correlated sources and use these sources with the common goal of efficiently computing a message. Similar to secret key capacity, a similar rate–capacity bound exists in DSC as well.

In secret sharing, one wishes to distribute a secret into several shares among users such that any $k$ out of $n$ users can all reconstruct the secret, whereas a subset smaller than $k$ would learn nothing at a reconstruction attempt. A common secret sharing scheme uses Reed-Solomon codes \cite{reed1960polynomial}, belonging to the class of maximum distance separable (MDS) codes. MDS codes under certain conditions can satisfy the secret sharing requirement: any $k$ out of $n$ shares suffice to reconstruct the secret, whereas every subset smaller than size $k$ is statistically independent of the secret. This optimal distance property means the scheme uses the smallest possible amount of redundancy for a given reconstruction threshold. These properties make Reed–Solomon and other MDS codes widely used for threshold secret sharing as well as in communication settings where error correction is necessary. Other works in coding theory explore security where no information is leaked for a subset of codewords in random linear codes \cite{cohen2018secure, cohen2021network, woo2023cermet, kim2018efficient}.

Recent works in information theory tie together MDS codes, MMI, and secret key agreement. Reing et al.\ explore properties of error-correcting codes maximizing MMI \cite{reing2019maximizing}. They derive a cohesion-$k$, a linear entropy measure similar to MMI that also quantifies the dependence of a $k$-variable subset for an $n$-variable system. Chan et al.\ \cite{chan2015multivariate} analyze multiterminal secret key capacity as an alternative measure of MMI.

\section{Preliminaries}
In this section we introduce
definitions that are used for the paper. We start with multiterminal communication.
\begin{definition}[Multiterminal communication]
    A terminal in our communication system that is used for secret key agreement can broadcast and receive messages from other terminals to communicate. Multiterminal communication uses multiple terminals that all communicate publicly to all other terminals.
\end{definition}

\begin{definition}[Operational SKA]
Consider a multiterminal setting with $V$ users (terminals), where each terminal can communicate with every other terminal over a public channel. Let there be $A \subseteq V$ active users and $h$ helping users such that $h = |V \setminus A|$. Active users must end up with the final secret key $K\in \mathcal{K}$, where $\mathcal{K}$ is the set of possible key values. Helpers need not end up with the secret key. All users can participate in the public discussion, over a noiseless, authenticated channel with the set of all public messages being $F$. Each user $i \in V$ has access to a private correlated source $Z_i$ and observes an $n$-sequence $Z^n_i$. Users broadcast some function of the private source over the public discussion channel that an eavesdropper can observe. The key must be uniform to the wiretapper
\[
\lim_{n\to\infty} \tfrac{1}{n}\,\bigl[\log|\mathcal{K}| - H(K \mid F)\bigr] \;=\; 0 \mbox{,}
\tag{1}
\]
where $H$ is the Shannon entropy. Lastly, the key must be recoverable by each active user in the sense that there exist functions $\phi_A = (\phi_i : i \in A)$
such that
\begin{equation}
  \lim_{n\to\infty}
  \Pr\!\bigl\{\,K \neq \phi_i\bigl(Z_i^{\,n},F\bigr)\bigr\}
  = 0,
  \qquad \mbox{  for all } i \in A.
  \tag{2}
\end{equation}
\end{definition}

\begin{definition}[Secret key capacity]
The secret key capacity is the maximum achievable key rate, a function of the active users and correlated sources
\begin{equation}
  C_{\mathrm S}(A, Z_V)
  = \sup_{K,F}
     \,\lim_{n\to\infty}\frac{1}{n}\,\log \lvert K\rvert,
  \tag{3}
\end{equation}
where the supremum is over the key and public discussion. 
\end{definition}

The secret key capacity can be calculated in terms of the communication for omniscience rate,  $R_{CO}$, for a set of active users, as proven in \cite[Theorem 1]{csiszar2004secrecy}: 
\[
C_{\mathrm S}(A, Z_V)
  = H(Z_V) - R_{CO}(A; Z_V).
\]
Next we define the basis of the SKA scheme we introduce in this work, error-correcting codes and specifically MDS codes. 
Consider a string $M$ of length $k$ over an alphabet $\Sigma$ with cardinality $|\Sigma| = q$. Error-correcting codes typically transform these strings into codewords of length $n \geq k$, and it is important to note the distance of such codewords $d = \min_{c_i,c_j \in \mathcal C} \triangle(c_i,c_j)$ where $\triangle$ is the Hamming distance, and $c_i,c_j \subset \mathcal C \subset q^n$. One is able to detect $k$ errors of Hamming distance one if $d \geq k+1$, and correct $k$ errors if $d \geq 2k+1$. A class within error-correcting codes is the class of linear codes. If we have a \textbf{linear code} with $n$-length, $k$-dimension, and $d$ distance over alphabet of size $q$, the code $\mathcal C= c_1,\dots,c_m, \mathcal C \subseteq \mathbb F_q^n$ is a $k$-dimensional subspace and therefore has exactly $m=|\mathcal C|=q^k$ codewords. 
\begin{definition}[MDS codes]
A subset of linear codes are within the \textbf{maximum distance separable (MDS) codes} class, if also $d = n - k+ 1$. 
\end{definition}
MDS codes can be represented by a basis $G = c_1, \ldots, c_k$ where each $c_i\in G$ has length $n$, and the span of $G$ produces all $q^k$ possible codewords. $G$ in this case is also called the generator matrix of $C$.

Within the class of MDS codes are Reed-Solomon (RS) codes. RS codes are often used in secret sharing schemes and other private linear encoding schemes due to the threshold property that any set of codewords less than $k$ reveals no information about an encoded string. Herein, consider a similar RS code, with the $k$-threshold security property as above. 
\begin{definition}[Reed-Solomon codes]\label{RS}
    We describe an $(n,k)$-RS code: fix a prime power $q$ and finite field $\mathbb{F}_q$ with $1\le k\le n\le q-1$. 
Fix non-zero pairwise distinct evaluation points $A=(\alpha_1,\ldots,\alpha_n)\in\Bbb F_q^n$, and non-zero column coefficients $v=(v_1,\ldots,v_n)\in \Bbb F_q^n$. Construct generator matrix, $G:$
\[
G(A,v)
=
\begin{bmatrix}
v_1 & v_2 & \cdots & v_n\\
v_1\alpha_1 & v_2\alpha_2 & \cdots & v_n\alpha_n\\
\vdots & \vdots & \ddots & \vdots\\
v_1\alpha_1^{k-1} & v_2\alpha_2^{k-1} & \cdots & v_n\alpha_n^{k-1}
\end{bmatrix},
\]
such that $G$ is full-rank. Also construct a string $m = (s,a_1,\ldots,a_{k-1})\in\Bbb F_q^k$ with a secret $s$, and uniform $a_1,\ldots,a_{k-1}$. The string is encoded as $c = mG = \bigl(v_1 f(\alpha_1),\ldots,v_n f(\alpha_n)\bigr)$, where
$
f(x)=s+\sum_{j=1}^{k-1} a_j x^j
$.
\end{definition}
Correctness and security (that no information is leaked with $k-1$ or below shares) is shown in \cite{mceliece1981sharing}. Other types of MDS codes  satisfy this security property \cite{cohen2018secure, tajeddine2018private, okada2002mds}, but for simplicity we just consider the class of RS codes.

MMI generalizes mutual information to measure dependencies among a set of $N$ random variables \cite{mcgill1954multivariate}. 
\begin{definition}[Multivariate mutual information]
    MMI can be expressed as 
    \[
I_{N}\!\bigl(X_{1};X_{2};\dots;X_{N}\bigr)
      = \sum_{k=1}^{N} (-1)^{k-1}
        \sum_{\substack{X \subset \{X_{1},X_{2},\dots,X_{N}\}\\ |X| = k}}
        H(X).
    \]
\end{definition}
A positive MMI indicates redundancy shared by every variable, whereas a negative value reveals more information together than the variables would separately.

    The use of error-correcting codes to maximize MMI has been explored in \cite{reing2019maximizing}, which derives cohesion-$k$, a linear entropy measure similar to MMI that also quantifies the dependence of a $k$-variable subset for an $n$-variable system. Chan et al.\ also analyze another form of MMI, \cite[Theorem 4.1]{chan2015multivariate} relating to multiterminal secret key capacity in the SKA case without helpers, solving for 
    \[
    C_{\mathrm S}(A, Z_V) = I(Z_V) 
    \]
    using the Dilworth truncation of residual entropy and the divergence bound of \cite{csiszar2004secrecy}.

\begin{definition}[Probabilistic polynomial-time adversary]
A probabilistic polynomial-time (PPT) adversary is an adversary whose computational power is bounded by a polynomial in the security parameter.     
\end{definition}

\begin{definition}[IND-CPA Security (e.g., see~\cite{KL14})]
An encryption scheme is said to be \emph{indistinguishable under chosen-plaintext attack} (IND-CPA secure) if no PPT adversary can distinguish encryptions of any
two equal-length messages of its choice with probability (non-negligibly) better
than $1/2$.
\end{definition}

Assuming integer factorization is hard (i.e., requires exponential time), public key cryptosystems achieve  computationally secure encryption (IND-CPA), so a PPT adversary is only able to extract negligible information.

\section{Code-based Secret Key Agreement Scheme}\label{scheme}
We now present our SKA scheme and prove its security.
Start by encoding a secret key with padded uniformity, $K = (s, u_1, \dots, u_{k-1})$ where s is the secret and $u_i$ are uniform pads, over a field $q$ using an $(n,k)$ RS code with generator matrix $G$, as in Def.~\ref{RS}, such that $\textbf{Z} = K G$. Each of the $|V|$ terminals, denoted by $Z_i$, obtains a single symbol (or partition) of the code $Z_i, i \in \{1, \ldots, |V|\}$. Next, broadcast $k-1$ symbols to every terminal so that every terminal has $k$ symbols that it can use to recover $K$. Because each terminal's correlated private source in SKA typically is not a symbol from an encoded secret key, we describe several settings where our MDS SKA scheme can be achieved and analyze each. Using an RS code, 
the following $n,k,q,$ and $G$ can be public information to an eavesdropper.

Start with a secret sharing inspired example: Let there be a main dealer terminal, $Z_1$, and $n$ total terminals. $Z_1$ establishes a preprocessed distinct shared (uniformly random and independent) value $r_i$ with each terminal $Z_i$, considered their correlated private sources. $Z_1$ takes the secret key and uniform pad, which we can call $K$, then encodes it with an $(n,k)$ code such that $\textbf{Z} = (z_1, \dots , z_n) = K G$. $Z_1$ can now either send each terminal a unique codeword $z_i$, or send the same codeword, say $z_1$, to all terminals. In both cases, this can be done by broadcasting $e_i = z_i + r_i$ for the unique codeword case or $e_i = z_1 + r_i$. Each terminal then receives their $e_i$ and subtracts their $r_i$ to recover their $z_i$ or $z_1$. $Z_1$ then broadcasts a subset of $k-1$ codewords that has not been sent to any terminal, $\mathcal C, \mathcal C \subset \textbf{Z}, |\mathcal C| = k-1$ and each party can reconstruct the secret key. The main benefit from sending each terminal unique codewords $z_i$ rather than $z_1$ is that the same message is not being distributed to all terminals. This reduces the chance of redundancy attacks, but comes at the cost of a larger $n$ and code block since we still need $k-1$ unique codewords for public discussion. To operate in the unique codeword setting, $n \ge |V| +k-1$, where $|V|$ is the number of terminals.

Alternatively, if there is no shared secret $r_i$ between the terminals and the main dealer terminal, one can use public key cryptography to distribute the shares and achieve computational security within the scheme.

A more general description of the scheme involves more than one codeword $z_1, \dots, z_i$ distributed to each terminal before public discussion using additional private sources or PKC. In this case, with $u = |z_1, \dots, z_i|$, $k-u$ codewords only are publicly discussed. We discuss necessity of this later. 

\begin{algorithm}[t]\label{mdsska}
\caption{MDS code-based SKA scheme}
\begin{algorithmic}
\State \textbf{Args:} $V$ terminals, $A \subset V$ active users, $(n,k)$ MDS code parameters
\State Dealer terminal generates $K$ and $(n,k)$ code generator matrix $G$
\State Dealer terminal computes $\textbf{Z} = K G$ and divides into codewords for each terminal
\For{$i = 1$ to $|V|$}
  \State Codeword distribution for each terminal $Z_i$
\EndFor
\State Broadcast $k-1$ unique codewords to $Z_1, \dots, Z_{|V|}$
\State Each terminal $Z_1, \dots, Z_{|V|}$ computes $K$
\end{algorithmic}
\end{algorithm}

Consider the cases where we use the unique private sources to distribute the code shares. In this case, even a computationally unbounded eavesdropper can only solve for the secret key with negligible probability.
\begin{theorem}
Suppose each terminal $Z_i$ receives a masked share $e_i = z_i + r_i$, where the masks $r_i$ are independent
uniform elements of a field of size $q$. Suppose the public discussion reveals exactly $k-1$ distinct
RS shares. Then the joint distribution of $(e_1, \ldots, e_n)$ and the publicly revealed shares are statistically independent of the secret $s$. 
Hence, Algorithm 1 leaks no information about the secret key to a computationally unbounded eavesdropper.
\end{theorem}
\begin{IEEEproof}
The construction of our scheme guarantees the security. Since the eavesdropper in the SKA setting does not know each $r_i$, it cannot decode the $e_i$ distributed. Subsequently, the eavesdropper only sees $k-1$ codewords, which are independent of the key. Because the last codeword that the eavesdropper sees is completely independent from their observed $e_i$, the $e_i$ and $k-1$ codewords (full transcript observed) are independent of the secret key, granting us information-theoretic security.
\end{IEEEproof}
The proof follows the assumption of an RS code-based secret sharing scheme \cite{mceliece1981sharing}. One issue that can arise is if we are using too small of a field $q$, an adversary may brute force the entire field and solve for the last codeword. Therefore it is important to choose an adequate field size; see Sec.~\ref{disc}.
Alternatively, when using PKC to distribute the codewords to each terminal, we retain computational security.

\begin{lemma}
    Suppose we use a computationally secure IND-CPA public-key encryption scheme to distribute $u$ shares to each terminal.
Then Algorithm 1 is computationally secure against PPT adversaries in our multiterminal SKA setting.
\end{lemma}
\begin{IEEEproof}
By contradiction: assume a similar setting from Theorem 1 with a PPT adversary that can decode the secret key. The adversary must have observed $k$ codewords. However, $k-u$ codewords are in the public transcript. 
If a PPT adversary could reconstruct the secret key, it would necessarily distinguish encryptions of different shares or recover plaintext shares without the secret key, contradicting IND-CPA security.
\end{IEEEproof}

\section{Secret Key Capacity for Code-based Key Agreement Scheme}

In this section, we derive meaningful capacity bounds for our MDS code-based scheme  in Sec.~\ref{scheme}; \cite{reing2019maximizing} shows that MMI is maximized for these codes. This expression gives the exact key rate attainable when every user participates, and it depends only on the code parameters $(n,k)$, and the alphabet size $q$. 

\begin{theorem}
For our secret key agreement scheme in Sec.~\ref{scheme}, based on an $(n,k)$ MDS code over $\mathbb{F}_q$, and supposing all terminals are active. Let there be $n$ terminals denoted by $Z_V = (Z_1, \dots, Z_n)$ such that each terminal has one symbol as their correlated private source post-share distribution. The achievable secret-key capacity is $C_{\mathrm S}(A, Z_V) = \frac{n-k}{n-1} \log q$.
\end{theorem}
\begin{IEEEproof}
Start with the secrecy capacity in (3). The quantity was upper-bounded by the following Csisz\'ar–Narayan divergence bound \cite{csiszar2004secrecy}:
\[
C_{\mathrm S}(A, Z_V)
\;\le\;
\min_{\substack{\mathcal P\in\Pi'(V):\\ C\cap A\neq\varnothing\;\forall\,C\in\mathcal P}}
\frac{1}{\lvert\mathcal P\rvert-1}\,
D\!\Bigl(
  P_{Z_V}\,\Big\|\,\prod_{C\in\mathcal P} P_{Z_C}
\Bigr)
\tag{4}
\]
where $\Pi'(V)$ is the set of partitions of $V$, and $D$ is the KL divergence. In the case without helpers, the secrecy capacity (3) is equivalent to the divergence bound (4) by modeling the bound in the form of MMI (4), see \cite{chan2015multivariate}. This leaves us with
\[
I(Z_V)
\;=\;
\min_{\mathcal P \in \Pi'(V)} I_{\mathcal P}(Z_V)
\]

\[
I_{\mathcal P}(Z_V)
\;=\;
\frac{1}{\lvert\mathcal P\rvert - 1}\,
D\!\Bigl(
  P_{Z_V}
  \,\Big\|\,
  \prod_{C \in \mathcal P} P_{Z_C}
\Bigr)
\]
\[
D\!\Bigl(
  P_{Z_V}
  \,\Big\|\,
  \prod_{C\in\mathcal P} P_{Z_C}
\Bigr)
\;=\;
\sum_{C\in\mathcal P} H(Z_C)\;-\;H(Z_V)
\tag{5}
\]
where \(\mathcal{P}\) is a partition of \(V\) into \(|\mathcal{P}| \ge 2\) parts, each of which must intersect \(A\). Since $V = A$, we use the result from \cite[Theorem 4.1]{chan2015multivariate} that MMI is equivalent to key capacity in the no-helper setting,
\[
I(Z_V) = C_{\mathrm S}(A, Z_V). \tag{6}
\]
With all the preliminaries, we now obtain the desired bounds for $(n,k)$ MDS codes. Consider multiparty $n$-user SKA, where all users are active (no helpers) and an $(n,k)$ code over a field of size $\mathbb{F}_q$. Each of the $n$ users receives symbol $z_i$ for $KG = \textbf Z = (z_1 ... z_n)$. In \cite{reing2019maximizing}, one uses the fact an MDS generator matrix $(G)$ is isomorphic to a uniform matroid, meaning that the entropy for the generator matrix, as well as an encoding using the matrix and a uniform message $(\textbf Z = KG)$ is
\[
H(Z_S) = \min(|S|, k) \log q,
\]
since the entropy is maximized with $k$ parts of an $(n,k)$ code ($S$ in this case is the number of codewords and recall that we need $k$ codewords in order to decode). Use 
\[
I_{\mathcal P}(Z_V)
\;=\;
\frac{1}{\lvert\mathcal P\rvert - 1}\,
\left(
  \sum_{C\in\mathcal P} H(Z_C)\;-\;H(Z_V)
\right)
\]
because each partition only has a singular symbol $z_i$, and the entropy of several partitions maximizes at $k \log q$,  
\[
\sum_{C\in\mathcal P} H(Z_C) = n \log q \text{ and } H(Z_V) = k \log q,
\]
yielding 
$
I_{\mathcal P}(Z_V) = \tfrac{n-k}{\lvert\mathcal P\rvert-1} \log q.
$
To satisfy (5), we use the singleton partition as ${\lvert\mathcal P\rvert - 1}$ decreases, but $n-k$ and $\log q$ remain constant, meaning ${\lvert\mathcal P\rvert = n}$,  giving us
\[
I(Z_V)
\;=\;
\min_{\mathcal P \in \Pi'(V)} I_{\mathcal P}(Z_V) = \tfrac{n-k}{n-1} \log q.
\]
This is also the secret key capacity according to (6). 
\end{IEEEproof}
Subsequently, we can use this 
result to also derive an upper bound of the capacity in the helper setting. 
\begin{prop}
Let $h$ be the number of helpers.
The key capacity can be upper-bounded by $\log q$ when $k\leq h+1$ and $\tfrac{n-k}{|A|-1}\log q$ when $k > h+1$.
\end{prop}
\begin{IEEEproof}
Note that with the existence of helpers, (5) is no longer tight due to the MMI becoming a relaxation of the omniscience dual when helpers are present, enabling us to only solve for an upper bound. Let there be $h = |V \setminus A|$ helpers. Start with (5):
\[
C_{\mathrm S}(A, Z_V)
\;\le\;
\tfrac{1}{\lvert\mathcal P\rvert - 1}\,
\left(
  \sum_{C\in\mathcal P} H(Z_C)\;-\;H(Z_V)
\right)
\]
From above, $\lvert\mathcal P\rvert =|A|$, since only the active users need to derive a key. Since $H(Z_C) = \min(|C|, k) \log q$, 
\[
\sum_{C\in\mathcal P} H(Z_C) = (|A|-1 + \min\{1+h, k\}) \log q,
\]
giving us the expression 
\[
I(Z_V)
\;=\;
\min_{\mathcal P \in \Pi'(V)} I_{\mathcal P}(Z_V) = \tfrac{|A|-1+k -k}{|A|-1} \log q = \log q,
\]
if $k \leq h + 1$; otherwise,
$C_{\mathrm S}(A, Z_V)
\;\le\;
\tfrac{n-k}{|A|-1} \log q.$
\end{IEEEproof}
\begin{prop}
The MDS codes used in our SKA scheme maximize the magnitude of McGill's MMI for an $(n,k)$-code over alphabet of size $q$.
\end{prop}
\begin{IEEEproof} To prove this, we calculate McGill's form of MMI \cite{mcgill1954multivariate} for our $(n,k)$-MDS code. Consider an $(n,k)$-code in field $\mathbb{F}_n^q$ with a full-rank generator matrix. We calculate the MMI for $n$ codewords. Start with
\[
    I_{N}\!\bigl(X_{1};\dots;X_{n}\bigr)
      = \sum_{i=1}^{n} (-1)^{i-1}
        \sum_{\substack{X \subset \{X_{1},X_{2},\dots,X_{n}\}\\ |X| = i}}
        H(X)
\]
and for our setting let $S$ be the subsets such that $\varnothing \neq S \subset \{Z_{1},\dots,Z_{n}\}$
\[
    I_{n}\!\bigl(Z_{1};\dots;Z_{n}\bigr)
      = \sum_{S}^{N} (-1)^{|S|-1}
    H(Z_S).
\]
Similar to above for a partition of codewords we have $H(Z_S) = \min(|S|,k)$, which yields the MMI as 
\begin{align*}
        I_{n}\!\bigl(Z_{1};\dots;Z_{n}\bigr)
      &= \sum_{S}^{n} (-1)^{|S|-1}
    \min(|S|,k) \log(q).\\
&= \Bigr( \sum_{i = 1}^n (-1)^{i-1} {n \choose i} \min(i,k)\Bigl) \log q\\
&= (-1)^{k-1} {n-2 \choose k-1} \log q.
\end{align*}
In the case where $n = k, {n-2 \choose k-1} =0,$ and $I_{n}\!(Z_{1};\dots;Z_{n}) = 0$. The MDS codes MMI value is either negative or positive depending on whether $k$ is even or odd, respectively. We can see that in both cases of MMI, the MDS code used in our scheme maximizes the magnitude of the classical MMI for the field size, as well as the key capacity. In \cite{reing2019maximizing}, their alternative derived cohesion MMI measure is also maximized for full-rank MDS codes.
\end{IEEEproof}

\section{Discussion}\label{disc}

In this section we discuss the results of our scheme considering parameter selections, limitations, and applications.

\subsection{Parameter selection}
We specify concrete parameter choices that make the code‑based SKA scheme resistant to brute force attacks for any PPT adversary. Let $u$ be the number of symbols each terminal begins with, and again let $k$ serve as the reconstruction threshold for the secret key. This means $k-u$ codewords are publicly distributed to an eavesdropper. The simplest brute-force attack from an adversary would be an attempt at decoding with every possible symbol combination, which has at minimum $q^u$ complexity. In practical settings, each terminal would have to start with hundreds of symbols.

\subsection{Applications}
We now discuss applicable settings where one is able to use the derived SKA scheme. We consider multiparty key refreshment and error correction in general key agreement. First consider the case of multiparty key refreshment, where we already have an established secret key and have multiple users. In practice, the use of the same symmetric key to encrypt a large amount of ciphers without refreshing into a new private key is not recommended. We can use $(n,k)$ MDS codes in the public discussion channels, given we only discuss a subset strictly less than $k$. As long as the key already being used is within the MDS codes' field, we can create a full MDS code with the existing private key. We can then decode the full codeword to obtain the new, refreshed private key. As before, the MI between the public discussion, what the eavesdropper observes, and the new secret key is zero since for any partition of the codeword less than $k$, the codeword is independent. 

Next, consider the case in which one is trying to establish a key over a noisy communication setting. Due to the inherent error-correcting property of MDS codes, we can forward additional codewords and still successfully establish the key. It is essential to ensure that any part of the codeword that fails to reach an intended
user is also unavailable to the adversary; otherwise, the adversary could obtain enough shares to reconstruct the secret.

\section{Conclusion}
We present an interesting characterization of MDS codes, specifically those used in secret sharing, but for secret key agreement. We then derive meaningful capacity bounds and MMI measures for our key agreement scheme, and discuss applications and implications for our scheme. Future work includes combining our scheme with encryption protocols such that it can be readily applied to real-world settings.

\bibliographystyle{IEEEtran}
\bibliography{references.bib}

\end{document}